# Identify the diapycnical eddy diffusivities by salt fingers and turbulence with vertical microstructure measurements

Liang SUN

Key Laboratory of Atmospheric Composition and Optical Radiation, CAS, School of Earth and Space Sciences, University of Science and Technology of China, Hefei, 230026, China.

**Abstract**

Diapycnical eddy diffusivities are formulated from physical relations according to a simple fact that the different formulas are identical for the same parameter. It is found that the dispassion ratio $\Gamma$ is a crucial parameter. When it is above a critical value (about 0.2), the flow is salt finger type; otherwise, it is turbulence. All the density ratio $R_\rho = 1/(1-\Gamma^2)$, eddy flux ratio $\gamma = 1/(1+\Gamma)$, and eddy diffusivity ratio $R_k = 1-\Gamma$ are simply dependent on the dispassion ratio $\Gamma$. We apply these relations to the measurement data in the western tropical Atlantic Ocean. The effective diapycnal diffusivity is $0.96 \times 10^{-4}$ m$^2$/s that agrees quite well with the observation from 0.8 to $0.9 \times 10^{-4}$ m$^2$/s by tracer.

## 1. Introduction

Much of the upper ocean Central Waters at tropical, subtropical and mid-latitudes are salt finger favourable (Kunze, 2003). The salt finger was observed by ocean investigations, e.g., the North Atlantic Tracer Release Experiment (NATRE, Ledwell et al., 1993), and the Caribbean Vorticity Experiment cruise (CaVortEx I, Morell et al., 2006). For growing salt-finger instabilities in the oceans, both background temperature and salinity must decrease with depth (Stern, 1960) and the density ratio $R_\rho$ must be less than the diffusivity ratio (~100). The effective diffusivities of salt and heat are formulated from microstructure measurements of the dissipation rates of turbulent kinetic energy ($\varepsilon$) and thermal variance ($\chi_\theta$), which are commonly obtained from dropped microstructure profilers (Schmitt et al., 2003). The data could be either explained by turbulence mixing or salt finger convection. Much depends on the methods of analysis used to interpret the available microstructure.

The role of salt fingers in oceanic mixing has been controversial. Some researcher emphasize the smallness of the net buoyancy flux (e.g., Gregg and Sanford, 1987). But others think they can provide diapycnal fluxes of heat, salt and density in the ocean thermocline more efficient than the turbulence does (e.g., Schmitt et al., 2005). It is hard to choose the right parameter to apply the formulas. For example, the effective diffusivity for heat was estimated at 0.17 to $0.45 \times 10^{-4}$ m$^2$/s by turbulence mixing or 1.03 to $2.4 \times 10^{-4}$ m$^2$/s by salt finger convection. However, the effective diffusivity for tracer and salt is 0.8 to $0.9 \times 10^{-4}$ m$^2$/s (Schmitt et al., 2005), which is just located in the middle of both results. It is hard to judge which one it is.

The goal of this study was to establish a simple formula to identify the effective diapycnal diffusivities of salt and heat by using the microstructure measurements of temperature and salinity profiles. We apply the results of previous studies, and combine them together according to a simple fact that the different formulas are identical for the same parameter.

## 2. Formulation

### 2.1 Basical definitions

The microstructure profilers provide the measurements of vertical temperature $T(z)$, salinity $S(z)$, and density $\rho(z)$ profiles along vertical axis *z*. The squared buoyancy frequency $N^2$ is

$$N^2 = -\frac{g}{\rho}\frac{\partial \rho}{\partial z} \tag{1}$$

where *g* is gravity. The density ratio $R_\rho$ is defined as below:

$$R_\rho = \alpha T_z / \beta S_z \tag{2}$$

where $\alpha$ is the thermal expansion coefficient, $\beta$ is the haline contraction coefficient, and the subscription "*z*" represents the vertical gradient. The eddy flux ratio $\gamma$ is defined as

$$\gamma = <\alpha T'w'> / <\beta S'w'> \tag{3}$$

where *w* is the vertical velocity, the supper primes represent the fluctuations. The dissipation rate of thermal variance $\chi_\theta$ and the dissipation rates of turbulent kinetic energy $\varepsilon$ can be obtained from dropped microstructure profilers.

### 2.2 Eddy diffusivities for salt finger

The diapycnal eddy diffusivity of heat (McDougall, 1988; Hamilton et al., 1989) is,

$$k_\theta = \frac{\chi_\theta}{2\theta_z^2}, \tag{4}$$

where $\theta_z$ is the mean vertical temperature gradient. This equation can be used for both turbulence and salt finger (Schmitt, 2003). However, the diapycnal eddy diffusivity of salt depends on the physics. For salt finger, the diapycnal eddy diffusivity of salt by using the dissipation rate of thermal variance is (Schmitt, 2003),

$$k_s = \frac{R_\rho}{\gamma} \frac{\chi_\theta}{2\theta_z^2}, \tag{5}$$

and the diapycnal eddy diffusivity of salt by using the dissipation rates of turbulent kinetic energy is (Schmitt, 2003),

$$k_s = \frac{R_\rho - 1}{1 - \gamma} \frac{\varepsilon}{N^2}. \tag{6}$$

Physically, the diapycnal eddy diffusivity of salt has only one value, if it is well defined. Thus Eq. (5) must identify to Eq. (6), this yields to

$$\frac{R_\rho - 1}{1 - \gamma} \frac{\varepsilon}{N^2} = \frac{R_\rho}{\gamma} \frac{\chi_\theta}{2\theta_z^2}. \tag{7}$$

If we define a new parameter "dissipation ratio" (Okay, 1985) as

$$\Gamma = \frac{\chi_\theta}{2\theta_z^2} \bigg/ \frac{\varepsilon}{N^2} = \frac{\chi_\theta N^2}{2\varepsilon \theta_z^2}. \tag{8}$$

Note that this dissipation ratio can be determined by the microstructure profilers. Then the dissipation ratio yields to

$$\Gamma = \frac{\chi_\theta N^2}{2\varepsilon \theta_z^2} = \frac{R_\rho - 1}{R_\rho} \frac{\gamma}{1 - \gamma}. \tag{9}$$

The eddy flux ratio can be determined by the density ratio (Kunze, 2003),

$$\gamma = R_\rho - \sqrt{R_\rho(R_\rho - 1)} \tag{10}$$

Substituting the above equation of (10) into (9), the density ratio yields to

$$R_\rho = \frac{1}{1-\Gamma^2}, \qquad (11)$$

and the flux ratio yields to,

$$\gamma = R_\rho - \sqrt{R_\rho(R_\rho-1)} = \frac{1}{1+\Gamma}. \qquad (12)$$

Then the eddy diffusivity ratio is

$$R_k = \frac{k_\theta}{k_S} = 1-\Gamma. \qquad (13)$$

The diapycnal eddy diffusivity of salt for salt could be obtained as,

$$k_s = \frac{R_\rho}{\gamma}\frac{\chi_\theta}{2\theta_z^2} = \frac{R_\rho}{\gamma}k_\theta = \frac{1}{1-\Gamma}k_\theta. \qquad (14)$$

As we can obtain $\theta_z$, $N^2$, $\varepsilon$ and $\chi_\theta$ from the measurements, we can determine the eddy diffusivity of heat by Eq. (4), the dissipation ratio by Eq. (8), and the diapycnal eddy diffusivity of salt by Eq. (14).

## 2.3 Eddy diffusivities for turbulence

In this case, the diapycnal eddy diffusivity of heat could be obtained by either Eq. (4) or by using $N^2$ and $\varepsilon$ (Osborn and Cox, 1972; Osborn, 1980),

$$k_\theta \leq 0.2\frac{\varepsilon}{N^2}. \qquad (15)$$

Substituting Eq. (4) to Eq. (14), and noting Eq.(9), it yields to,

$$\Gamma \leq 0.2. \qquad (16)$$

The above inequality implies that the dissipation ratio is an index of mixing type. If $\Gamma \leq 0.2$, the mixing is turbulence, otherwise the mixing is salt finger, as shown in Fig. 1. Other parameters (the density ratio, the eddy flux ratio, and the eddy diffusivity ratio) can be calculated from the dissipation ratio. The above relations are shown in Fig. 1. In the observation, $\Gamma$ is expected to be within 0.4 to 0.9 (St Laurent & Schmitt, 1999).

## 2.4 Application

Tracer release experiments have been successfully carried out in an eastern subtropical gyre area (the North Atlantic Tracer Release Experiment, NATRE). From the high-resolution profiler survey, overall averages of $\varepsilon$ and $\chi_\theta$, the vertical temperature and density gradients were formed between the potential density anomaly. The effective diapycnal diffusivity for heat was estimated at 0.17 to $0.45 \times 10^{-4}$ m$^2$/s by turbulence mixing or 1.03 to $2.4 \times 10^{-4}$ m$^2$/s by salt finger convection (Schmitt et al., 2005).

Now, we use the same data to find that eddy diffusivity of heat $k_\theta$ is $0.45 \times 10^{-4}$ m$^2$/s, the dissipation ratio $\Gamma = 0.53$, and effective diffusivity of salt $k_S$ is $0.96 \times 10^{-4}$ m$^2$/s. This value is very close to the effective diapycnal diffusivity for tracer and salt of 0.8 to $0.9 \times 10^{-4}$ m$^2$/s (Schmitt et al., 2005).

The formulas can also be applied to other ocean observations, e.g. CaVortEx I (Morell et al., 2006). They are supposed to be valuable for the studies in the ocean mixing.

## 3. Conclusions

In this study, the diapycnal eddy diffusivities are formulated from the physical relations. It is found that the dispassion ratio $\Gamma$ is a crucial parameter. When it is above a critical value (about 0.2), the flow is salt finger type; otherwise, it is turbulence. All the density ratio $R_\rho = 1/(1-\Gamma^2)$, flux ratio $\gamma = 1/(1+\Gamma)$, and eddy diffusivity ratio $R_k = 1-\Gamma$ are simply dependent on the dispassion ratio $\Gamma$. And the diapycnal eddy diffusivities can be obtained from the high-resolution profiler measurements.

## Acknowledgements

We thank Prof. Zhou S.Q. and Prof. W. Wang for their discussions. This work was supported by the National Basic Research Program of China (No. 2012CB417402) and the National Natural Science Foundation of China (No. 41376017).

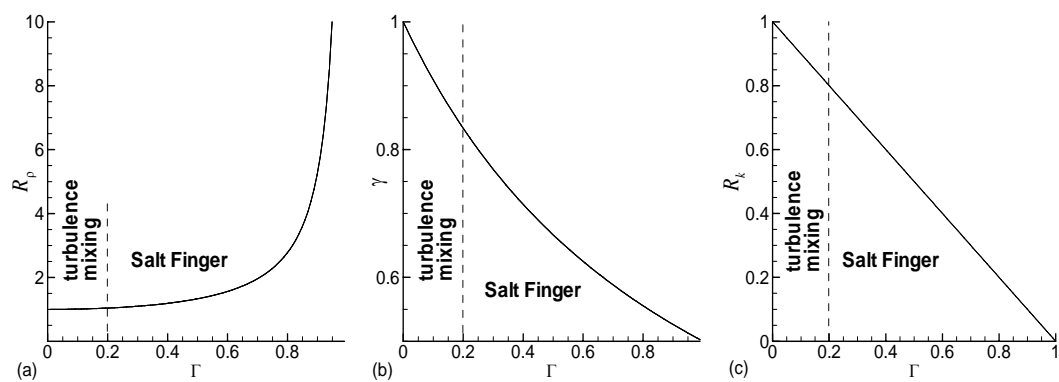

Figure 1. (a) Density ratio vs eddy dissipation ratio. (b) Eddy flux ratio vs eddy dissipation ratio. (c) Eddy diffusivity ratio vs eddy dissipation ratio.